\documentclass[twocolumn,10pt,prd,nofootinbib,floatfix,superscriptaddress,preprintnumbers]{revtex4-1}
\usepackage{placeins}
\usepackage{soul}
\usepackage[utf8]{inputenc}
\usepackage{sidecap}

\usepackage{hyperref}
\usepackage[dvips]{graphicx}
\usepackage{epsfig,amsmath,amssymb,verbatim,mathrsfs,array,layout,textcomp,amssymb,latexsym,slashed,graphicx,booktabs,mathtools}
\usepackage{cleveref}
\usepackage[dvipsnames]{xcolor}

\newcommand{\beq}{\begin{eqnarray}}
\newcommand{\eeq}{\end{eqnarray}}

\def\beqa{\begin{eqnarray}}
\def\eeqa{\end{eqnarray}}
\newcommand{\no}{\nonumber}
\newcommand{\bv}{\left(\begin{array}{c}}
\newcommand{\ev}{\end{array}\right)}
\newcommand{\bmtwo}{\left(\begin{array}{cc}}
\newcommand{\bmthree}{\left(\begin{array}{ccc}}
\newcommand{\emn}{\end{array}\right)}
\newcommand{\bmtwoc}{\left\{\begin{array}{cc}}
\newcommand{\bmthreec}{\left\{\begin{array}{ccc}}
\newcommand{\emnc}{\end{array}\right\}}
\newcommand{\ba}{\begin{array}}
\newcommand{\ea}{\end{array}}

\newcommand{\avg}[1]{\left< #1 \right>}
\newcommand{\abs}[1]{\left| #1 \right|}

\newcommand{\dbar}{d\hspace*{-0.08em}\bar{}\hspace*{0.1em}\hspace{0.05cm}}

\newcommand\numberthis{\addtocounter{equation}{1}\tag{\theequation}}

\newcommand{\Tr}{{\text{Tr }}}

\def\lsim{\mathrel{\rlap{\lower4pt\hbox{\hskip1pt$\sim$}}
     \raise1pt\hbox{$<$}}}         
\def\gsim{\mathrel{\rlap{\lower4pt\hbox{\hskip1pt$\sim$}}
     \raise1pt\hbox{$>$}}}    

\begin{document}

\title{
SIMPs through the axion portal
}

\author{Yonit Hochberg}\email{yonit.hochberg@mail.huji.ac.il} \affiliation{Racah Institute of Physics, Hebrew University of Jerusalem, Jerusalem 91904, Israel}
\author{Eric Kuflik}\email{eric.kuflik@mail.huji.ac.il} \affiliation{Racah Institute of Physics, Hebrew University of Jerusalem, Jerusalem 91904, Israel}
\author{Robert McGehee}\email{robertmcgehee@berkeley.edu} \affiliation{Department of Physics, University of California, Berkeley, CA 94720, USA}
\affiliation{Ernest Orlando Lawrence Berkeley National Laboratory, University of California, Berkeley, CA 94720, USA}
\author{Hitoshi Murayama}\email{hitoshi@berkeley.edu, hitoshi.murayama@ipmu.jp} \affiliation{Department of Physics, University of California, Berkeley, CA 94720, USA}
\affiliation{Ernest Orlando Lawrence Berkeley National Laboratory, University of California, Berkeley, CA 94720, USA}
\affiliation{Kavli Institute for the Physics and Mathematics of the
  Universe (WPI), 
  University of Tokyo, Kashiwa 277-8583, Japan}
\affiliation{DESY, Notkestra\ss e 85, D-22607 Hamburg, Germany}
\author{Katelin Schutz}\email{kschutz@berkeley.edu} \affiliation{Department of Physics, University of California, Berkeley, CA 94720, USA}
\affiliation{Ernest Orlando Lawrence Berkeley National Laboratory, University of California, Berkeley, CA 94720, USA}

\preprint{DESY 18-101, IPMU18-0114}

\begin{abstract} \noindent
Dark matter could be a thermal relic comprised of strongly interacting massive particles (SIMPs), where $3\to2$ interactions set the relic abundance. Such interactions generically arise in theories of chiral symmetry breaking via the Wess-Zumino-Witten term. In this work, we show that an axion-like particle can successfully maintain kinetic equilibrium between the dark matter and the visible sector, allowing the requisite entropy transfer that is crucial for SIMPs to be a cold dark matter candidate. Constraints on this scenario arise from beam dump and collider experiments, from the cosmic microwave background, and from supernovae. We find a viable parameter space when the axion-like particle is close in mass to the SIMP dark matter, with strong-scale masses of order a few hundred MeV. Many planned experiments are set to probe the parameter space in the near future.
  \end{abstract}

\maketitle

\section{Introduction}
Dark matter (DM) comprises the majority of the matter budget of the Universe, but its microphysical properties and origin remain unknown. One possibility is that DM is a thermal relic from the early Universe. The most well-studied thermal scenario is that DM is comprised of weakly interacting massive particles (WIMPs). The number density of WIMPs is set by $2\to2$
annihilations of the DM into Standard Model (SM) particles, and the observed DM relic abundance is achieved when both the DM mass and coupling to SM particles are near the scales relevant for electroweak processes.

An alternative thermal setup was proposed in Ref.~\cite{Hochberg:2014dra} where $3\to2$ DM self-interactions set its 
abundance. In this scenario, the observed relic 
density indicates that the DM mass and self-coupling should be near the strong scale. This mechanism of strongly interacting massive
particles (SIMPs) was shown to be generic in strongly coupled theories of chiral symmetry breaking, where the pions play the role of
DM~\cite{Hochberg:2014kqa}. The $3\to2$ interactions are then sourced by the well known Wess-Zumino-Witten (WZW) action~\cite{Wess:1971yu,Witten:1983tx,Witten:1983tw}. This provides a simple and calculable realization of the SIMP mechanism, although by no means the only one \cite{Lee:2015uva,Bernal:2015xba,Hochberg:2015vrg,Lee:2015gsa,Choi:2016hid,Bernal:2017mqb,Choi:2017mkk,Hochberg:2017khi,Choi:2017zww,Berlin:2018tvf,Choi:2018iit}. 

In addition to providing a novel thermal mechanism for explaining the dark matter abundance, SIMPs also offer a possible explanation for issues related to small-scale structure formation. In particular, observed dark matter subhalos tend to be less dense than in simulations (see Ref.~\cite{Bullock:2017xww} for a recent review). While many of these issues may be resolved with better understanding of astrophysical processes (for instance in Ref.~\cite{Wetzel:2016wro}), it is also possible to mitigate these issues if the dark matter can self-scatter (see Ref.~\cite{Tulin:2017ara} for a recent review). The strong self-annihilations of SIMP dark matter imply that their self-scatterings are also large, such that they naturally address these small-scale puzzles~\cite{Hochberg:2014dra,Hochberg:2014kqa}. 

The $3\to 2$ DM annihilations would raise the temperature of the residual DM due to conservation of comoving entropy. 
Therefore, 
the DM must be in thermal equilibrium with a heat sink, such as the SM bath, until after freeze-out~\cite{Hochberg:2014dra}.
Otherwise, the $3\to 2$ DM annihilations would cause the steady depletion of DM particles and heating of the remaining DM, a scenario referred to as cannibalization. While cannibalization was originally proposed to provide a class of DM models intermediate between hot and cold DM, such models are not observationally-viable~\cite{Carlson:1992fn,deLaix:1995vi}. Obtaining the observed DM abundance inevitably leads to an unacceptable washout of small-scale structure.

To allow for adequate thermalization between the DM and the SM, Refs.~\cite{Hochberg:2015vrg,Lee:2015gsa} explored the kinetically-mixed hidden photon portal. 
Here, we explore the possibility of a pseudoscalar portal using axion-like particles to accomplish the entropy transfer to photons. 
For brevity, we refer to axion-like particles simply as ``axions''    throughout the paper. 
We note that Ref.~\cite{Kamada:2017tsq} also considered an axion portal, but focused on the regime where semi-annihilations set the relic abundance. 
In contrast, we focus on the SIMP regime where $3\to2$ annihilations determine the relic density. 
For concreteness, we will use the SIMPlest pion realization of the DM based
on an Sp$(2N_c)$ gauge theory with four doublet Weyl fermions following Ref.~\cite{Hochberg:2014kqa}. Sp$(2N_c)$ gauge groups with a larger number of flavors or SU$(N_c)$ and SO$(N_c)$ gauge groups allow for semi-annihilations which can control the relic abundance, although there may still be parameter space where $3\to2$ annihilations determine the dark matter density. 

This article is organized as follows. In Section~\ref{sec:not}, we 
describe the framework and identify the interactions responsible for  
setting the correct DM relic abundance and cooling the DM via the axion portal. In order to cool the DM effectively, the axion must be in thermal equilibrium with both the DM and the SM.
In Section~\ref{sec:pionALP}, we 
illustrate the theoretical and empirical requirements 
of axion-pion thermal equilibrium, while in Section~\ref{sec:ALPSM} we do the same for axion-SM thermal equilibrium. 
Concluding remarks and discussions follow in Section~\ref{sec:conclusions}. 

\section{The framework}\label{sec:not}
Our starting point is an Sp(2$N_c$) gauge theory with $2N_f$ Weyl fermions that couple to an axion-like field $a$ as
\beq\label{eq:Laq}
 {\cal L}_{aq}=-\frac{1}{2} m_a^2 a^2 -\left( \frac{1}{2} m_Q e^{ia/f_{a\pi}}
 J^{ij} q_i q_j + {\rm h.c.} \right)
\eeq
where $m_a$ is the axion mass, $m_Q$ is the quark mass matrix, $q_i$ are the confining quarks and $J$ is the Sp($2 N_f$) group invariant.\footnote{Note that in terms of a 4-component spinor $\psi^T=(q
\quad q^\dagger)$, the identities $i\bar \psi \gamma_5 \psi=-i qq+i q^\dagger q^\dagger$ and $\bar \psi \psi=qq+q^\dagger q^\dagger$ hold.} Upon dynamical chiral symmetry breaking, the ground state is expected to be given by
\beq
\langle q_i q_j \rangle = \mu^3 J_{i j}.
\eeq
Any transformation by the flavor symmetry $V \in \text{SU}(N_f)$ would also be a ground state, and 
in general
\beq
\langle q_i q_j \rangle = \mu^3 (VJV^T)_{i j}.
\eeq
Switching the description to the chiral Lagrangian, a spacetime-dependent flavor rotation gives the low-energy excitations,
\beq
\langle q q \rangle \rightarrow \mu^3 \Sigma\,,\quad \Sigma \equiv V J V^T\,, \quad V=\exp(i \pi/ f_\pi)\,,\
\eeq
where $\pi\equiv \pi^b T^b$, $T^b$ are the Sp($2 N_f$) generators and $f_\pi$ is the pion decay constant. We use the normalization $\text{Tr }T^b T^c = 2 \delta^{bc}$ for the generators.
In terms of the pion fields,
\beq\label{eq:qtopi}
\begin{array}{rcl}
-\frac{i}{2} m_Q J^{ij} q_i q_j + h.c. \quad &\Rightarrow& \quad \frac{m_\pi^2}{6 f_\pi}  {\rm Tr}\;  \pi^3 + \mathcal{O}(\pi^5)\,, \\
- \frac{1}{2} m_Q J^{ij} q_i q_j + h.c. \quad &\Rightarrow&  \quad -\frac{m_\pi^2}{4}  {\rm Tr}\;  \pi^2 + \mathcal{O}(\pi^4)\,.
\end{array}
\eeq
The theory has an SU($2N_f$)/Sp($2N_f$) flavor structure, where the residual Sp($2N_f$) is exact due to the quark masses' proportionality to $J$. 
For $N_f\geq 2$, the fifth homotopy group of the coset space is non-vanishing and the WZW term exists~\cite{Wess:1971yu,Witten:1983tx,Witten:1983tw},
\begin{equation}
  {\cal S}_{\rm WZW}
  = \frac{-i N_c}{240\pi^2} \int \Tr (\Sigma^\dagger d \Sigma)^5\, .
\end{equation}

Generally, both $a{\rm Tr}\left(\pi^3\right)$ and $a^2{\rm Tr}\left(\pi^2\right)$ terms can appear 
in the interaction Lagrangian. However, the former introduces semi-annihilations of pions into a pion and an axion which might contribute to determining the relic abundance of the dark matter. Here, we are interested in exploring the role of an axion mediator in the SIMP mechanism of $3\to2$ self-annihilations of pions.
Consequently, we focus on an Sp$(2N_c)$ gauge theory with $N_f=2$ fermions, where the flavor symmetry is SU(4)/Sp(4) and $N_\pi = (N_f-1)(2N_f+1)=5$ pions emerge. In this theory, the semi-annihilation process is absent since ${\rm Tr}\left(\pi^3\right)=0$, and pure $3\to2$ annihilations of pions via the WZW term are guaranteed to control the relic abundance of DM. For more flavors, or for other gauge groups, $3\to2$ annihilation may still control the relic abundance, though in a smaller region of parameter space.
To leading order in pion fields, the WZW term for our choice of gauge group takes the form
\begin{eqnarray}\label{eq:wzw}
  \lefteqn {{\cal L}_{\rm WZW}
  = \frac{2 N_c}{15 \pi^2 f_\pi^5}\epsilon^{\mu\nu\rho\sigma} \Tr\!\!\left[\pi \partial_\mu
  \pi \partial_\nu \pi \partial_\rho \pi \partial_\sigma \pi \right] } \nonumber \\
  &=& \frac{8 N_c}{15 \pi^2 f_\pi^5}\epsilon^{\mu\nu\rho\sigma} \epsilon_{abcde}
  \pi^a \partial_\mu \pi^b \partial_\nu \pi^c \partial_\rho \pi^d \partial_\sigma \pi^e.
\end{eqnarray}

The excess kinetic energy generated in the dark sector from $3\to2$ annihilations needs to be transferred out, which can be obtained through kinetic coupling of the pions to the axions and the axions to the SM bath. Since the semi-annihilation term is absent for our flavor group of choice, the interaction Lagrangian between pions and axions is 
\beq\label{eq:Lapi}
{\cal L}_{a\pi} \supset \frac{\kappa}{4} a^2 \pi^b \pi^c \delta^{bc}\,.
\eeq
If the axion coupling to the pions arises in a similar manner to what occurs in QCD, as in Eq.~\eqref{eq:Laq}, the mass term for the hidden quarks $q$ in
the Sp(2$N_c$) gauge theory gives rise to an axion potential:
\begin{align*}
\label{eq:Laq2}
{\cal L}_{a\pi}&= -\frac{1}{2} m_a^2 a^2 -\frac{1}{2}m_Q \mu^3  e^{i a/f_{a\pi}} {\rm Tr}J \Sigma +{\rm h.c.}\\ &= -\frac{1}{2} \left( m_a^2 a^2 +\frac{2 m_\pi^2 f_\pi^2}{f_{a\pi}^2} \right) a^2
+\frac{m_\pi^2}{8f_{a\pi}^2}a^2 {\rm Tr}\pi^2+\cdots \numberthis
\end{align*} 
where $m_\pi$ is the pion mass.
Using the normalization of ${\rm Tr} \pi^2 = 2 \pi^b \pi^c \delta^{bc}$, we
identify the Feynman rule for the $a a \pi^b \pi^c$ vertex in Eq.~\eqref{eq:Lapi} as
\beq\label{eq:lam}
i \kappa \delta^{bc} = i \frac{m_\pi^2}{f_{a\pi}^2} \delta^{bc} \,.
\eeq
Meanwhile, the interaction Lagrangian between the
axions and SM photons is 
\beq\label{eq:Laphoton}
{\cal L}_{a\gamma}&=&\frac{1}{4 f_{a\gamma}} a F^{\mu\nu} \tilde F_{\mu\nu}\ .
\eeq

  As long as the kinetic equilibrium between the pions and the SM is maintained through the interactions of Eqs.~\eqref{eq:Laq2} and~\eqref{eq:Laphoton}, the preferred mass for the dark matter is $m_\pi \approx 300$~MeV \cite{Hochberg:2014kqa} with $m_\pi \sim 2\pi f_\pi$ to set the observed relic abundance. We find below that viable ALP masses are around the same scale, $10\text{ MeV}\lsim m_a \lsim 1$~GeV. Couplings that satisfy $m_\pi \sim 2\pi f_\pi$ correspond to the strongly-interacting regime of the theory, where self-interactions are important on astrophysical scales. In this regime, ${\cal O}(1)$ corrections to perturbative results are expected, and therefore should be thought of as a proxy for the scales involved. Phenomenologically interesting pion masses lie at the edge of perturbativity, where higher order corrections and vector meson effects can impact the range of observationally-viable pion masses~\cite{Hansen:2015yaa,Choi:2018iit}.

\section{Axions and pions in equilibrium} 
\label{sec:pionALP}
\subsection{Theoretical requirements}
The interaction Lagrangian in Eq.~\eqref{eq:Laq2} leads to annihilations of pions into axions and to elastic scattering of axions off of pions. The SIMP mechanism requires the former process to be suppressed at the time of $3\to 2$ freeze-out while the latter is active. 

The requirement in order for the $3\to2$ pion self-annihilations to dominate  
the $2\to2$ annihilations of pions into axions 
at the time of freeze-out is  
\beq\label{eq:annsupp}
n_\pi\langle\sigma v\rangle_{\rm ann} \lsim H|_{T_F}\,,
\eeq
where $\langle\sigma v\rangle_{\rm ann}$ is the thermally averaged cross section for the annihilation process $\pi\pi \to aa$.
The Hubble parameter at freeze-out is given by \beq H|_{T_F}=\sqrt{\frac{g_{*,F} \,\pi^2 }{90}}\,\frac{T_F^2}{M_{\rm Pl}},\eeq where $T_F$ is the freeze-out temperature of $3\to2$ interactions (typically $T_F\sim m_\pi/20$ in the SIMP setup), $M_{\rm Pl}$ is the Planck mass, and $g_{*,F}$ is the effective number of relativistic degrees of freedom at the time of freeze-out. We have verified numerically that this requirement on the annihilation rate does maintain the correct relic abundance as set by the $3\to2$ SIMP mechanism.

The thermally averaged annihilation cross section that appears in Eq.~\eqref{eq:annsupp} can be readily calculated. For a trivial matrix element, ${\cal M}$, of a process $12\to 34$ (as is relevant for the Lagrangian of Eq.~\eqref{eq:Laq2})  
in which all states obey Maxwell-Boltzmann statistics, the thermally averaged cross section entering the Boltzmann equations~\cite{Gondolo:1990dk} is
expressed in terms of  
\begin{align}\label{eq:gammafunc}
\gamma_{12 \rightarrow 34} = & \, \frac{g_1 g_2  g_3 g_4 T |\mathcal{M}|^2}{2^9 \pi^5}\int_{s_{\min}}^\infty ds \,\sqrt{s} \,\lambda^{1/2}(\sqrt{s}, m_1,
m_2)\no\\
&\times 
\lambda^{1/2}(\sqrt{s}, m_3, m_4) K_1({\sqrt{s}}/{T})\,, 
\end{align}
where $g_i$ counts degrees of freedom for particle $i$, \beq s_{\min} = {\max\{(m_1+m_2)^2,(m_3+m_4)^2\}}, \eeq  \beq \lambda(x, y, z) \equiv (1-(y+z)^2/x^2)(1-(y-z)^2/x^2),\quad \eeq and $K_{1}$ is the first order modified Bessel function of the second kind. The amplitude $|\mathcal{M}|^2$ which appears is averaged over \emph{all} degrees of freedom. For pion-axion scattering and pion annihilation to axions, the relevant amplitude is therefore 
\beq 
\abs{\mathcal{M}}^2 \equiv
\frac{1}{g_\pi} \frac{1}{g_\pi} \sum_{b,c} \kappa^2 \delta^{bc} = 
 \frac{m_\pi^4}{g_\pi f_{a \pi}^4}
 \label{matrix}
 \eeq
since the trace requires that the pions be the same.
The thermally averaged cross
section for the annihilation process $\pi\pi \to aa$ is
\beq\label{eq:xsecgen}
\langle \sigma v \rangle_{\rm ann} &=& \frac{1}{2} \frac{\gamma_{\rm ann}}{(n_{\pi}^{\rm eq})^2}\,,
\eeq
where $n_i^{\rm eq}$ denotes the number density of particle $i$ in equilibrium,
\beq\label{eq:neqNR}
n_i^{\rm eq}=\frac{g_i}{2\pi^2}T_i m_i^2 K_2(m_i/T_i)\,,
\eeq
where $K_{2}$ is the second order modified Bessel function of the second kind. In Eq.~\eqref{eq:xsecgen}, the  
phase-space factor of $1/2$ for identical initial particles  
is cancelled because the number density changes by 
two particles per annihilation. %
For $m_a\ll m_\pi$,
Eq.~\eqref{eq:xsecgen} simplifies to
\beq\label{eq:xsecsim}
\langle \sigma v\rangle_{\rm ann}&\approx&\frac{g_a^2m_\pi^2}{64 g_\pi \pi f_{a \pi}^4}\,.
\eeq

In addition to the suppression of the $2\to 2$ annihilations, the SIMP mechanism requires that the rate of energy transfer in the scattering process $a\pi\to a\pi$  
is fast enough to successfully cool the DM. 
We require that thermal decoupling occurs after freeze-out, $T_D<T_F$,  
so that the energy transfer is efficient for the entirety of the freeze-out process. In the limit of small $m_a$ (using Bose-Einstein statistics), we follow the analytic derivation of Ref.~\cite{Kuflik:2017iqs} which obtains the thermal decoupling temperature $T_D$ of the $\pi a \to \pi a$ elastic scattering process \beq T_D \sim m_\pi \left( \frac{\pi g_a^2 m_\pi^5}{120  f_{a \pi}^4 H|_{T=m_\pi}}\right)^{-1/4}, \quad \quad  m_a \ll T_F. \quad \quad\eeq The details of this derivation, including a normalization factor of $4^{-1/4}\Gamma(3/4)$, can be found in Ref.~\cite{Kuflik:2017iqs}. We have checked numerically that this requirement on thermalization between the axions and pions does keep the DM cool.

In the regime where $T_F<m_a \lesssim m_\pi$, we generalize the approach laid out in Ref.~\cite{Bringmann:2006mu} for particles scattering in the limit of low momentum transfer. We estimate that in the viable parameter space, this approximation is valid to within 10\% even though the pion mass is larger than the axion mass by only order unity factors. Working to second order in the momentum transfer and using Maxwell-Boltzmann statistics for the axion, we find that decoupling occurs when 
\beq \frac{3 g_a^2  m_\pi m_a^2\, T_D^2~ e^{-m_a/T_D}}{ (2 \pi)^3 f_{a \pi}^4 H|_{T=m_\pi}}  \sim 1,~~ T_F<m_a \lesssim m_\pi . ~~~~~
\eeq
We find that the low- and high-mass axion decoupling temperatures match onto each other when numerical differences between Bose-Einstein and Maxwell-Boltzmann statistics are taken into account in the intermediate regime. The details of the derivation can be found in Appendix~\ref{app1}. 

In addition to the above requirements, the decay constant $f_{a\pi}$ 
must be greater than the cutoff scale of chiral symmetry breaking.
Otherwise, the description in Eq.~\eqref{eq:Laq} breaks down. 
We require that $f_{a\pi}\gsim 2\pi f_\pi$,
where $f_{\pi}$ is determined for a given $m_\pi$ from the solution to the Boltzmann equation. Since the Sp(2$N_c$) gauge theory with $N_f=2$ we discuss here
points to the strongly interacting regime where $m_\pi \sim 2\pi f_\pi$, we require that $f_{a\pi}\gsim m_\pi$. 
In practice, however, suppressing $2\rightarrow 2$ annihilations at freeze-out is always a stronger requirement. 

An additional preference, though not a requirement, comes from considering how chiral symmetry breaking contributes to the axion mass in Eq.~\eqref{eq:Laq2}, 
\beq
\Delta m_a^2=\frac{2 m_\pi^2 f_\pi^2}{f_{a\pi}^2}\,.
\eeq
The natural range for the axion mass-squared is therefore where $\Delta m_a^2 \lesssim m_a^2$, such that no fine tuning is present against an unspecified negative contribution, possibly from another confining gauge theory with $\theta \approx \pi$.

\begin{figure}[t!]
\begin{center}
\includegraphics[width =.49 \textwidth]{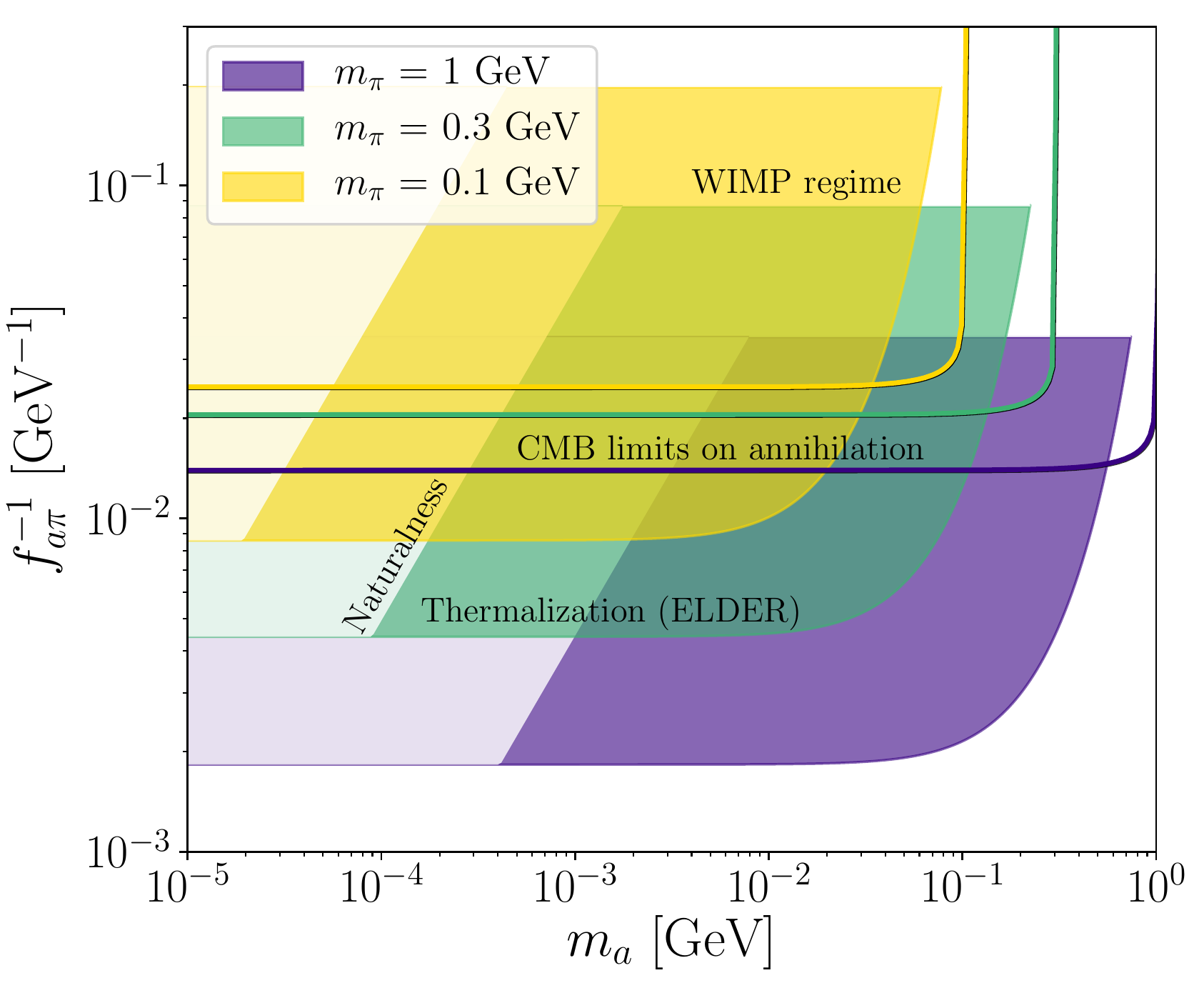}
\caption{\label{fig:pionALP}
The parameter space for axions coupling to pions. The shaded regions correspond to the regions where the SIMP mechanism is theoretically viable for a given dark matter mass $m_\pi$. The boundaries of this region are set by requiring that the $3\to2$ rather than the $2 \to 2$ process sets the relic abundance (labeled as ``WIMP regime'') and  that the pions can transfer their excess heat to the axions and hence the SM sector (labeled as ``Thermalization''). Note that the parameter space along the ``Thermalization'' curve corresponds to the scenario where dark matter is an elastically decoupled relic (ELDER). We also indicate the natural mass range where the axion mass is at least as large as its contribution from chiral symmetry breaking (labeled as ``Naturalness''). Also shown are the empirical upper limits on pion annihilation from energy injection into the CMB (thick solid lines labeled as ``CMB limits on annihilation'').
}\end{center}
\end{figure}

Satisfying the above requirements on $f_{a\pi}$ as a function of $m_a$ for a variety of dark matter masses $m_\pi$ yields the viable SIMP regions depicted in Fig.~\ref{fig:pionALP}. 
We take $g_{*,F}=10.75$ at freeze-out since for the DM masses we consider, freeze-out happens below the temperature of muon-antimuon annihilation. 
We learn that viable SIMP-axion thermalization is achieved over a broad range of axion masses and couplings $f_{a\pi}$. 

We note that elastically decoupling relic (ELDER) dark matter~\cite{Kuflik:2015isi,Kuflik:2017iqs} is obtained along the thermalization curve in Fig.~\ref{fig:pionALP}. For ELDER DM, the kinetic decoupling between the DM and SM baths occurs before $3\to2$ pion self-annihilations freeze out. This causes the relic abundance to be exponentially sensitive to this elastic scattering while being relatively insensitive to the strength of the $3\to2$ pion process. On the thermalization curve in Fig.~\ref{fig:pionALP}, the elastic scattering of pions off of axions dominates over the $3\to2$ pion self-annihilation process in setting the relic abundance.

\subsection{Empirical requirements}
Having established the theoretical requirements on the axion-SIMP parameter space, we now move to the observational constraints coming from the cosmic microwave background (CMB). 
In standard cold dark matter cosmology, the intergalactic medium (IGM) is almost entirely neutral after recombination and CMB photons free stream. If some fraction of the DM annihilates to SM particles and partially ionizes the IGM, this will cause some CMB photons to re-scatter which modifies CMB anisotropies in a characteristic way. For the scenario we consider in this work, the process of interest is $ \pi \pi \to a a \to 4 \gamma$. 
In the parameter space where there is sufficient thermalization between axions and the SM (see Section~\ref{sec:ALPSM}), the decay of the intermediate-state  
axions happens immediately. Thus, the use of the narrow-width approximation is appropriate and the cross section for this process is set by the cross section for the annihilation process $\pi \pi \to a a$. We use limits derived in Ref.~\cite{Slatyer:2015jla}, which are not very sensitive to
whether there are two final-state photons or four~\cite{convo}. 
The resulting upper limits 
are shown in Fig.~\ref{fig:pionALP} as a set of thick, solid lines corresponding to the different depicted pion masses. 
We thus find a viable SIMP-axion parameter space below the CMB curve and above the thermalization curve.

\section{Axions and photons in equilibrium}
\label{sec:ALPSM}
%

\subsection{Theoretical requirements}

Fig.~\ref{fig:pionALP} presents the viable region where the SIMP and axion maintain thermal equilibrium. If the axion and SM also maintain thermal equilibrium via the axion-photon coupling in Eq.~\eqref{eq:Laphoton}, then the pions will share a temperature with the SM.
For most of the axion masses we consider, decays and inverse decays into SM photons are the most efficient processes for kinetic equilibrium with the SM at freeze-out. 
 The rate for these decays at rest is 
\beq
\Gamma_{a} = \frac{m_a^3}{64\pi f_{a\gamma}^2}\, .
\eeq
For the axions to thermalize with the SM during  freeze-out, two conditions must be satisfied. First, axion decays and inverse decays must be fast enough to thermalize the axions with the SM, \beq\label{eq:aSMkin}
\frac{\Gamma_{a} m_a e^{-m_a/2 T_F}}{4 T_F H|_{T_F}} &\gsim& 1\,.
\eeq This is the strongest condition in the regime where the axion is relatively light and abundant. 
In the regime where the axion mass is comparable to or larger than that of the pion, a second condition on their decay becomes stronger than just pure thermalization between the axions and SM:
\beq\label{eq:heavyALPtherm} \frac{\Gamma_a T_F^2 n_a}{m_a} \gtrsim \frac{H|_{T_F} m_\pi^2 n_\pi}{T_F}. \eeq   
This second condition requires that the axions decay quickly enough to transfer entropy that has accumulated from the pion sector to the SM. 
This matters more for higher axion masses $m_a \gtrsim m_\pi$
since the axion number density is lower than the pion number density, 
so that each axion decay must transfer several pions' entropy. 
The detailed derivations of these conditions can be found in Appendix~\ref{app2}.

Decays and inverse decays come into equilibrium at late times. \emph{A priori}, this could suggest
the need for Eqs.~\eqref{eq:aSMkin} and~\eqref{eq:heavyALPtherm} to hold prior to freeze-out in order to sufficiently transfer entropy from the annihilating pions into the SM particles. However, we verified numerically that this is not the case: the SIMP relic abundance is unaffected if decays and inverse decays into SM particles only come into equilibrium close to the time of freeze-out.

\begin{figure*}[t!]
\includegraphics[width =\textwidth]{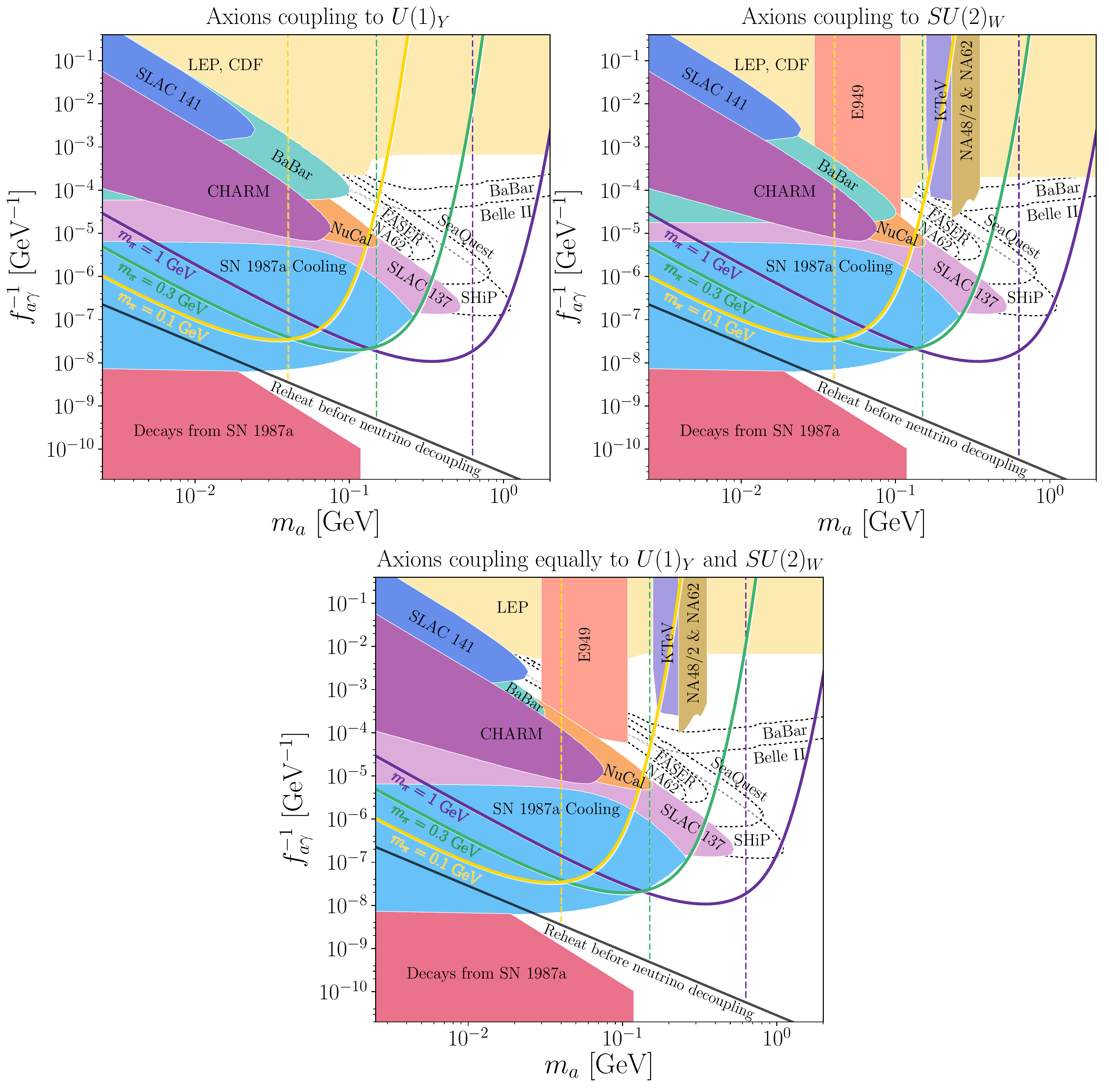}
\vspace{-0.3cm}
\caption{\label{fig:aGbound}
The parameter space for axions coupling to photons through U$(1)_Y$ (top left), SU$(2)_W$ (top right), and both in equal amounts (bottom). 
Solid curves correspond to the lower bound on the decay rate for thermalization between the axions and SM for various pion masses. 
Vertical, dashed lines correspond to the largest axion mass allowed by CMB constraints for a given pion mass (see Fig.~\ref{fig:pionALP}). 
Below the thermalization curves, the SIMP mechanism may still be viable down to the black solid line if the axion decays out of equilibrium and reheats above the neutrino decoupling temperature. 
Solid, filled regions correspond to existing constraints from supernova 1987a \cite{Dolan:2017osp, Jaeckel:2017tud}, LEP and CDF \cite{Acciarri:1994gb,Jaeckel:2015jla,Dolan:2017osp}, BaBar \cite{Dolan:2017osp}, beam dump experiments SLAC 137, SLAC 141, CHARM, and NuCal~\cite{Dolan:2017osp,Bjorken:1988as,Riordan:1987aw,Hewett:2012ns,Dobrich:2015jyk,Dobrich:2017gcm}, and kaon decay experiments E949, NA62, NA48/2, and KTeV~\cite{Artamonov:2005ru,Ceccucci:2014oza,Abouzaid:2008xm, Izaguirre:2016dfi}. Regions enclosed by dotted, black lines correspond to projected reach by SHiP~\cite{Alekhin:2015byh,Dobrich:2015jyk}, NA62~\cite{Dobrich:2015jyk}, Belle II 3$\gamma$~\cite{Izaguirre:2016dfi,Dolan:2017osp} (noting that the projected Belle II constraint from $\gamma + \text{invisible}$ falls between NuCal and NA62~\cite{Dolan:2017osp}), BaBar~\cite{Izaguirre:2016dfi},  SeaQuest~\cite{Berlin:2018pwi} and FASER~\cite{Feng:2018pew}. See the main text for more details.
 } \vspace{-0.3cm}
\end{figure*}

For axions with masses below the freeze-out temperature, the scattering process $ae\to \gamma e$ is more efficient than the decays considered above. This arises because the rate of scattering is enhanced relative to decays like $\sim(T/m_a)^3$ for axions that are relativistic at the time of freeze-out, while the rate of decays is suppressed due to the boost factor of $\sim(m_a/T)$. 
We find that the parameter space with $m_a<T_F$ is tightly constrained for the pion masses we consider and therefore do not include the scattering process $ae\to \gamma e$ in our analytics or numerics, since it is expected to be subdominant for axion masses with $m_a\gtrsim T_F$. Note that by including decays as the only channel to transfer entropy, we are being conservative since adding the $ae\to \gamma e$ channel would only lower the required coupling strength between axions and the SM.

In Fig.~\ref{fig:aGbound}, we depict the requirement on $f_{a\gamma}$ such that decays and inverse decays
sufficiently transfer entropy between the sectors.
Each solid curve corresponds to the lower bound on $f_{a\gamma}^{-1}$ to maintain thermal contact for a fixed pion mass. 
We use the full Boltzmann equations and full energy transfer rates.
A crossover between two regimes occurs at $m_a\sim m_\pi$, where the lower axion number density starts to matter and 
Eq.~\eqref{eq:heavyALPtherm} becomes a stronger condition than Eq.~\eqref{eq:aSMkin}.
In the regime where $m_a\gtrsim m_\pi$, kinetic equilibrium is maintained by axions in the exponential tail of the distribution, which causes the precipitous increase in $f_{a\gamma}^{-1}$.
As is evident, kinetic equilibrium between the axion and the 
SM through decays and inverse decays is possible over a range of axion masses.

The conditions outlined above amount to requiring that the axions and SM have the same temperature at freeze-out; however, there is another possibility which we outline here. For the DM to cool, it only needs to transfer entropy to the axions throughout the  freeze-out process. Then, instead of keeping the axions and SM at the same temperature, axions can decay out of equilibrium into the SM at some later time. Relative to the scenario where the axions are thermalized with the SM, the pion--axion sector will be slightly hotter than the SM and the relic abundance will be slighter larger for the same value of $3\to2$ rate. Therefore if the axions and SM are not thermalized at freeze-out, the value of $f_\pi$ must be increased slightly to give the right relic abundance.
For sufficiently large $m_a$, the universe undergoes a brief matter-dominated phase where the axions dominate the energy density of the universe. When the axions decay, they reheat the SM components, the universe becomes radiation-dominated again, and the pion abundance is diluted. This happens when 
\beq
H(T_\text{RH}) =\sqrt{\frac{g_{*, \text{RH}} \pi^2 }{90}} \frac{T_\text{RH}^2}{M_{\rm Pl}} \sim \Gamma_a.
\eeq
We require that the reheat temperature be larger than the temperature of neutrino decoupling: if the reheat temperature is lower, then the photons get preferentially heated and the effective number of relativistic neutrinos ($N_\text{eff}$) becomes smaller than allowed by observations of the CMB~\cite{planck}. We take the neutrino decoupling temperature to be $\sim$3~MeV~\cite{enqvist1992neutrino}, at which point $g_* = 10.75$.
Having such a high reheat temperature also enforces that the decay products do not affect Big Bang Nucleosynthesis (BBN). 
Therefore, the region in Figs.~\ref{fig:aGbound} between the solid curves and straight line labeled ``Reheat before neutrino decoupling'' may be viable with a slightly different $3\to2$ cross-section than in the standard SIMP scenario. More dedicated numerical studies are necessary in this case and will be explored in future work.

\subsection{Empirical requirements}
Having established a theoretically-viable parameter space, we must check whether it is allowed by current experiments and observations.
Constraints arise from early universe cosmology, astrophysical bodies, and terrestrial experiments.
\subsubsection{Light Degrees of Freedom}
If light axions are in thermal equilibrium with the SM bath, a bound on their mass arises from their effect on the temperature ratio
$T_\nu/T_\gamma$ after neutrinos have decoupled. This difference alters the effective number of neutrino species contributing to the radiation density, $N_{\rm eff}$, which can be measured in the CMB by comparing the photon diffusion scale to the sound horizon scale~\cite{planck,Boehm:2013jpa}. Such constraints are relevant for light particles in equilibrium with the photon 
or electron plasma beneath the temperature of neutrino
decoupling unless the particle couples to neutrinos as well. When applicable, this bound is stronger than the BBN bound of
$\sim$~MeV, which comes from the fact that changes to $N_\text{eff}$ change the expansion history and hence modify the abundance of the light elements. Because of this bound, only values of $m_a > 2.6$~MeV are shown Fig.~\ref{fig:aGbound}.

\subsubsection{Supernova 1987a}

The direct coupling of the axion to 
photons can lead to excess emission from supernovae (SN) via the Primakoff scattering
process~\cite{Masso:1995tw}. When the coupling between axions and the SM is sufficiently strong, the scattering of $e\gamma \to e a$ produces axions in the stellar medium which leads to
excess cooling if the axions escape the SN. However, if the coupling is too strong, then trapping occurs via the inverse $ea \to e\gamma$ process along with axion decays, in which case the axion does not carry any energy out of the star. SN cooling primarily proceeds through neutrino emission;
due to the observed neutrino signal from SN 1987A, any new SN cooling process must carry away less energy than the neutrinos, $\sim 3 \times 10^{53}$ ergs.
The region of parameter space excluded by the excess cooling of SN 1987A~\cite{Dolan:2017osp} is shown in Fig.~\ref{fig:aGbound}. 
For photon couplings that are too weak to produce significant energy loss in the supernova, there are still constraints from escaping axions
decaying into an observable burst of photons~\cite{Jaeckel:2017tud}, which we also show in Fig~\ref{fig:aGbound}.

\subsubsection{Terrestrial}

The couplings between axions and SM particles are constrained by terrestrial experiments. However, these constraints often come with assumptions about how the axions interact with the SM.
We classify constraints on the axion-photon coupling based on different assumptions about its fundamental origin, namely that the photon coupling arises from: 
\begin{enumerate}
\item solely coupling to $U(1)_Y$; 
 \item solely coupling to $SU(2)_W$;
\item equal couplings to $U(1)_Y$ and $SU(2)_W$, in which case, the $aZ\gamma$ coupling vanishes. 
 \end{enumerate} 

Measurements from the LEP collider and CDF constrain the decay $Z\to \gamma \gamma (\gamma)$~\cite{Acciarri:1994gb,Jaeckel:2015jla,Dolan:2017osp} as shown in Fig.~\ref{fig:aGbound}. BaBar also constrains the decay $Z\to \gamma + \text{invisible}$~\cite{Dolan:2017osp}. 
In the third case above, in which the $aZ\gamma$ coupling vanishes due to equal couplings to $U(1)_Y$ and $SU(2)_W$, both of these $Z$ decay constraints are alleviated.
In their place, there is a LEP bound on $e^+ e^- \to \gamma \gamma$~\cite{Jaeckel:2015jla} and a BaBar bound on $e^+ e^- \to \gamma +\text{inv}$~\cite{Dolan:2017osp}.
Constraints from electron beam dump experiments SLAC 137, SLAC 141~\cite{Dolan:2017osp,Bjorken:1988as,Riordan:1987aw,Dobrich:2017gcm}, 
and proton beam dump experiments CHARM and NuCal~\cite{Hewett:2012ns,Dobrich:2015jyk} apply for axions coupled to photons regardless of how the coupling arises. 
Constraints from $K_L\to \pi^0 a$ and $K^{\pm}\to \pi^{\pm}a$ with $a\to \gamma\gamma$ assume the axion couples to $SU(2)_W$. These kaon results were obtained in Ref.~\cite{Izaguirre:2016dfi} from analyses of fixed-target kaon rare decay experiments by the E949~\cite{Artamonov:2005ru}, NA62 and NA48/2~\cite{Ceccucci:2014oza}, and KTeV~\cite{Abouzaid:2008xm} collaborations. 

In addition to existing constraints, we show the projected reach of several future experiments and analyses on the photon coupling to an axion, indicated by the dashed black curves in Fig.~\ref{fig:aGbound}.
We include the projected reach of SHiP~\cite{Alekhin:2015byh,Dobrich:2015jyk}, NA62~\cite{Dobrich:2015jyk}, BaBar~\cite{Izaguirre:2016dfi}, Belle II~\cite{Izaguirre:2016dfi,Dolan:2017osp}, SeaQuest~\cite{Berlin:2018pwi} and FASER~\cite{Feng:2018pew}. In principle there could be a constraint from a process involving the $a ZZ $ coupling for all three scenarios, though we expect it would be weaker than the constraints we present and at this time we are not aware of any existing or projected limits from such a process. 

\section{Discussion}\label{sec:conclusions}

In this paper, we have considered the pion realization of SIMP dark matter in strongly coupled gauge theories, and have shown that it can be realized with axions as the thermalization portal between the dark matter and SM. Throughout this work, we have required that all three sectors --- the SIMPs, axions, and SM --- share the same temperature as the $3\to2$ annihilations freeze out. This requirement sets a target range of masses and couplings for this mechanism to be theoretically viable. 

In examining the couplings between the SIMPs and axions, we have required that the coupling is strong enough to thermalize the two sectors via 2$\to$2 scattering. At the same time, we require that the coupling not be strong enough for 2$\to$2 annihilations to overwhelm the 3$\to$2 process that is the hallmark of the SIMP mechanism. Combined, these requirements lead to a well-defined range of couplings between the pion dark matter and the axions such that the SIMP mechanism can work. Constraints on annihilation coming from the CMB narrow the allowed range of couplings, though a broad parameter space remains. It is possible that a future CMB spectral distortions experiment can probe this parameter space further, though exploring this possibility is beyond the scope of this work.

Considering the couplings between the axions and the SM, we focused on the coupling to photons. 
For a given pion mass, there is a range of axion masses allowed for maintaining SIMP-axion equilibrium. Within this axion mass range, the main requirement for axion-SM thermalization is that the axions decay quickly enough to successfully transfer the entropy from the pions to the SM, which can easily be achieved. 
The relevant couplings to photons can be probed in a multitude of ways. The range of axion masses considered here are at an energy scale that is relevant for supernovae, which constrains part of the parameter space. Additionally, terrestrial beam dump and collider experiments have probed complementary parameter space. We find that the SIMP mechanism can be realized in a broad region of parameter space that is not excluded by current constraints. Several upcoming experiments are forecast to probe much of the viable parameter space that is currently allowed, providing an excellent handle for testing the framework. 

There are several possible ways to extend the parameter space for axion-mediated SIMPs. 
Some of these possibilities are already excluded by existing limits. For instance, heavy axions which mediate the entropy transfer through off-shell interactions (both through the interactions we consider here and through the CP-violating interaction $\mathcal{L} \supset \frac{f_{\text{CPV}}}{2} a \pi^b \pi^b$) are excluded by the LEP constraint shown in Fig.~\ref{fig:aGbound}. Another heavily-constrained scenario is that the axions couple to all fermions through a universal Yukawa coupling, which is almost entirely ruled out by SLAC 137, CHARM, kaon decays, B decays, supernova 1987a, BaBar and the muon anomalous magnetic moment~\cite{Essig:2010gu,Dolan:2014ska,Bauer:2017ris}. 
A more promising possibility is that axions couple only to electrons or to charged leptons; in this case, there are known limits from SLAC 137 and the muon anomalous magnetic moment~\cite{Essig:2010gu,Bauer:2017ris}.  
In addition, the axion-electron parameter space should be constrained by limits from supernova 1987a and from loop-induced couplings to photons. 
However, such constraints have not been explored in the literature and will be the subject of future work~\cite{future}. 
Another possibility is that the axions have a long enough lifetime that they decay out of equilibrium, dumping entropy to the SM and diluting the SIMPs--- this too will be explored in future work~\cite{future}.

\section*{Acknowledgments}
  We thank Babette Dobrich, Eder Izaguirre, Tongyan Lin and Tracy R. Slatyer for useful conversations and correspondence pertaining to this work. YH, EK and HM also thank Tomer Volansky and Jay Wacker for early collaboration on this project in 2014. We acknowledge the importance of equity and inclusion in this work and are committed to advancing such principles in our scientific communities.
YH is supported by
the Israel Science Foundation (grant No. 1112/17), by the Binational Science Foundation (grant No. 2016155), by the I-CORE Program of
the Planning Budgeting Committee (grant
No. 1937/12), by the German Israel Foundation (grant No. I-2487-303.7/2017), and  by the Azrieli Foundation. EK is supported by the Israel Science Foundation (grant No.
1111/17), by the Binational Science Foundation  (grant No. 2016153) and by the I-CORE Program of
the Planning Budgeting Committee (grant
No. 1937/12). 
HM is supported by the U.S. DOE under Contract DE-AC02-05CH11231, and
by the NSF under grants PHY-1316783 and PHY-1638509. HM is also
supported by the JSPS Grant-in-Aid for Scientific Research (C)
(No.~26400241 and 17K05409), MEXT Grant-in-Aid for Scientific Research
on Innovative Areas (No. 15H05887, 15K21733), by WPI, MEXT, Japan, by the Binational Science Foundation (No. 2016155), and the Alexander von Humboldt Foundation.
RM and KS are supported by the National Science Foundation Graduate Research Fellowship Program. KS is also supported by a Hertz Foundation Fellowship.\\

\appendix

\section{Boltzmann Equations}
\label{app:rev}
The Boltzmann equations govern the evolution of the phase space $f_X(p, t)$ for particle $X$, \beq \frac{\partial f_X}{\partial t} - H \frac{p^2}{E}\frac{\partial f_X}{\partial E} = C[f_X]\eeq
where the left hand side is the relativistic Liouville operator in a Friedmann-Robertson-Walker spacetime and $C[f_X]$ is the collision term. In the regime where $2 \to 2$ annihilations are negligible, the relevant terms which appear in the collision term are the $3 \to 2$ interactions which set the relic abundance, the DM self-interactions which give it a thermal distribution, the elastic interactions which transfer energy between the pions and axions, and decays (and inverse decays) of axions to SM particles. We neglect axions converting to photons (and vice versa) via $t$-channel scattering off electrons, which is less efficient than decays (and inverse decays) at thermalizing the axions with the SM.
In the parameter space of interest (with the exception of axions in the out-of-equilibrium decay scenario), all particles will be interacting sufficiently frequently so that they have thermal distributions, 
\beq f_X = \frac{1}{e^{(E-\mu)/T} \pm 1},\eeq where the $+$ sign ($-$ sign) is for Fermi-Dirac (Bose-Einstein) statistics and $\mu$ is the chemical potential. 
At temperatures below the mass of a given particle, the effects of quantum statistics become negligible and the Maxwell-Boltzmann (MB) distribution is recovered.

To solve the Boltzmann equations, it is most useful to look at two moments of the phase space distribution, which correspond to the number density and energy density \begin{align} n_X &= g_X \int \dbar^3 p \,f_X\\ \rho_X &= g_X  \int \dbar^3 p\, E \,f_X, \end{align}
where $g_X$ is the number of degrees of freedom and $\dbar^3 p \equiv d^3p/(2\pi)^3$.
The Boltzmann equations for axion-mediated SIMPs are \onecolumngrid
\beqa 
\frac{\partial n_{\pi}}{\partial t } + 3 H n_{\pi} &=& - \langle \sigma_{3 \to 2} v^2\rangle \left( n_{\pi}^3 - n_{\pi}^2 n_{\pi}^{\rm eq, T_\pi} \right), \label{BE1} \\
\frac{\partial \rho_{\pi}}{\partial t} + 3 H \left(\rho_{\pi} + p_{\pi}  \right) &=& -\langle \sigma_{el} v \delta E\rangle n_ {\pi} n_{a},  \\
\frac{\partial n_{a}}{\partial t } + 3 H n_{a} &=& - \left( \left< \Gamma\right>_{T_a} n_{a} - \left< \Gamma \right>_{T_{\rm SM}} n_{a}^{\rm eq, T_{SM} }\right),  \\ 
\frac{\partial \rho_{a}}{\partial t} + 3 H \left(\rho_{a} + p_{a}  \right) &=&  \langle \sigma_{el} v \,\Delta E\rangle n_{\pi} n_{a} -m_a \Gamma \left( n_{a} - n_{a}^{\rm eq,T_{SM }} \right). 
\eeqa
where $p_X$ is the pressure densities of species $X$ (which is related to the energy density through the equation of state $w_X$), and $n_{X}^{\rm eq, T}$ is the thermal equilibrium density for species $X$ at temperature $T$. Additionally, $\left< \Gamma\right>_{T} = \Gamma \left< m_a/E_a \right>_T$ is the thermally averaged decay rate of the axion at temperature $T$, $\langle \sigma_{3 \to 2} v^2\rangle$ is the thermally-averaged $3\to2$ cross-section of the pions for this choice of gauge group, labeled $i=1 \ldots 5$,
\begin{align} 
\langle \sigma_{3 \to 2} v^2\rangle &= \frac{1}{3! 2! n_\pi^3} \int \left( \prod_{i=1}^5 \frac{\dbar^3 p_i}{2 E_i}\right) f_1 f_2 f_3 \abs{\mathcal{M}_{123\to45}}^2(2 \pi)^4 \delta^4(p_1+p_2+p_3 -p_4-p_5),\\
& = \frac{6 N_c^2}{\sqrt{5}\pi^5}\frac{m_\pi^3 T_F^2}{ f_\pi^{10}}
\end{align}
and $n_{\pi} n_{a} \langle \sigma_{el} v \delta E\rangle$ is the energy transfer rate between the pions and axions (with the initial and final states labeled as $1$ and $2$),
 \beq n_\pi n_a \langle \sigma_{el} v\, \Delta E\rangle = \int \frac{\dbar^3 p_{\pi_1}}{2 E_{\pi_1}}\frac{\dbar^3 p_{a_1}}{2 E_{a_1}}\frac{\dbar^3 p_{\pi_2}}{2 E_{\pi_2}}\frac{\dbar^3 p_{a_2}}{2 E_{a_2}} (E_{\pi_1} - E_{\pi_2}) f_{\pi_1} f_{a_1} \abs{\mathcal{M}_{\pi a\to \pi a}}^2 (2 \pi)^4 \delta^4(p_{\pi_1}+p_{a_1} -p_{\pi_2}-p_{a_2}).~~~~\eeq
For MB statistics, the equilibrium values for the number, energy, and pressure densities for a particle of mass $m$ and temperature $T$ with $g$ degrees of freedom are 
\begin{align}
n^{\rm eq} &= g\int \dbar^3 p \,f^{\rm eq}  =  \frac{g m^2 T}{2\pi^2}K_2\left(\frac{m}{T}\right)\\
\rho^{\rm eq} &= g\int \dbar^3 p\, E f^{\rm eq}  =  \frac{g m^2 T}{2\pi^2}\left( m K_1\left(\frac{m}{T}\right) + 3T K_2\left(\frac{m}{T}\right)\right) \\
p^{\rm eq} &= g\int \dbar^3 p \frac{p^2}{3 E} f^{\rm eq}  =  \frac{g m^2 T^{ 2}}{2\pi^2}K_2\left(\frac{m}{T}\right)
\end{align}
and the thermally-averaged boost factor is 
\beq
\left< \frac{m}{E} \right>_{T} =\frac{1}{n^{\rm eq}}\int g \frac{d^3 p}{(2\pi)^3} \frac{m}{E}f^{\rm eq}  = \frac{K_1(m/T)}{K_2(m/T)}.~~ \label{thermboost}
\eeq
\section{Pion-axion kinetic decoupling}\label{app1}
We require that the pions and axions are in kinetic (thermal) equilibrium during the entire time over which the $3\to2$ process is active. This can be recast as a requirement that the thermal decoupling temperature between the two sectors is lower than the temperature at which the $3\to2$ process freezes out. In this range of temperatures, the pions are guaranteed to be nonrelativistic since $T_F \sim m_\pi/20$. Therefore, the energy transfer rate can be re-written as 
\beq  n_\pi n_a \langle \sigma_{el} v\, \Delta E\rangle \simeq - \int \frac{\dbar^3 p_{\pi_1}}{2 E_{\pi_1}} \frac{p_{\pi_1}^2}{2 m_\pi} C[f_{\pi_1}]. \label{eq:transferrate}\eeq
The form of the collision term in the integrand will depend on whether the axions are relativistic or not at around the time of freeze-out, as detailed below.
\subsection{Relativistic axions}
In the regime where the axions are still relativistic at the time of freeze-out, there are well known methods for computing the collision term \cite{Bringmann:2006mu}. For pion-axion scattering in this regime, the collision term takes the form 
\beq C[f_{\pi_1}] =  \frac{ \pi  g_a^2 m_\pi^6  }{360 f_{a \pi}^4} \left(\frac{T_a}{m_\pi}\right)^4 \left( m_\pi T_a\, \nabla^2_{p_{\pi_1}} + \vec{p}_{\pi_1}\cdot \vec{\nabla}_{p_{\pi_1}} +3 \right) f_{\pi_1}(T_\pi).\eeq
Integrating over the pion phase space in Eq.~\eqref{eq:transferrate} yields 
\beq
n_\pi n_a \langle \sigma_{el} v \,\Delta E\rangle = \frac{ \pi g_a^2 m_\pi^5 }{120 f_{a \pi}^4} \left(\frac{T_a}{m_\pi}\right)^4 (T_\pi - T_a) . \label{eq:elscatterterm}
\eeq
While the $3 \to 2$ is actively depleting the number density, the pions are nonrelativistic and follow MB statistics, which means that their energy density Boltzmann equation can then be expressed as 
\beq 
\frac{\partial T_\pi}{\partial T_a} = 3 \frac{T_\pi^2}{m_\pi T_a} + \frac{ \pi g_a^2  m_\pi^5 }{120 f_{a \pi}^4 H|_{T=m_\pi}}   \frac{ T_a T_\pi^2 (T_\pi - T_a)}{m_\pi^4}. \label{eq:equalization}
\eeq
The first term on the right-hand side comes from the $3\to 2$ and causes the pion temperature to increase, while the second term comes from pion-axion elastic scattering and pushes $T_\pi \to T_a$. The second term cannot keep up with the first as the temperature drops and the pions and axions decouple. The temperature of decoupling is 
\beq 
\label{eq:nonrel_answer}
T_D \simeq m_\pi \left(\frac{ \pi g_a^2  m_\pi^5 }{120 f_{a \pi}^4 H|_{T=m_\pi}}  \right)^{-1/4}. 
\eeq

\subsection{Non-relativistic axions}
The standard result for the collision term derived in Ref.~\cite{Bringmann:2006mu} applies only when the axion is relativistic. However, as long as the momentum transfer is still small, we can still apply the same methods as Ref.~\cite{Bringmann:2006mu} in deriving the collision term. We will be interested in the regime where the axion mass is still smaller than the pion mass (which kinematically enforces that the momentum transferred in a single collision is relatively small) but where the axion is sufficiently heavier than the freeze-out temperature $T_F\sim m_\pi/20$ such that it becomes Boltzmann suppressed. Most generally, the collision term for $2\to2$ scattering of pions and axions can be written as
\onecolumngrid
 \begin{align}  C = \frac{1}{2} \int \frac{\dbar^3 p_{a_1}}{2 E_{a_1}} \frac{\dbar^3 p_{a_2}}{2 E_{a_2}} \frac{\dbar^3 p_{\pi_2}}{2 E_{\pi_2}} (2 \pi)^4 \delta^{(4)}(p_{a_1}+p_{\pi_1} -p_{a_2} -p_{\pi_2}) \abs{\mathcal{M}}^2 J\end{align}
where $J$ is the relevant combination of phase space factors. In the regime of interest, everything is MB distributed at thermal decoupling so \beq J = e^{-E_{\pi_1}/T_\pi} e^{-E_{a_1}/T_a} - e^{-E_{\pi_2}/T_\pi} e^{-E_{a_2}/T_a}. \eeq
The collision term can be written as an expansion in the momentum transfer, $C = \sum C^j$ where
\beq C^j = \frac{(2 \pi)^4}{2 j!} \int \frac{\dbar^3 p_{a_1}}{2 E_{a_1}} \frac{\dbar^3 p_{a_2}}{2 E_{a_2}} \frac{\dbar^3 p_{\pi_2}}{2 E_{\pi_2}}  \delta(E_{a_1} + E_{\pi_1} - E_{a_2} -E_{\pi_2}) \abs{\mathcal{M}}^2 J \big( (\vec{p}_{a_2} - \vec{p}_{a_1})\cdot \vec{\nabla}_{p_{\pi_2}}\big)^j \delta^{(3)}(p_{\pi_1} - p_{\pi_2}).\quad \eeq
In this expansion, $C^0$ vanishes simply because if the momentum transfer is zero and the number of a species does not change, the collision term is identically zero. For a contact interaction which has no angular dependence (as in the scenario we consider here,  $\abs{\mathcal{M}}^2\sim$~const.), $C^1$ also vanishes because the angular integral contains the integrand $(\vec{p_{a_2}} - \vec{p_{a_1}})\cdot \vec{p_\pi}\equiv \vec{q} \cdot \vec{p}_{\pi_1}$, which is odd over the angular domain. Therefore, the leading-order term is $C^2$. The momentum transfer scales like $\Delta p_a\sim (m_\pi p_a - m_a p_\pi)/(m_a+m_\pi)$, and plugging in thermal values for the typical momentum indicates that when truncated at $\mathcal{O}((\Delta p_a/p_\pi)^2)$, the expansion in is accurate at the $\sim$10\% level when the axion mass is $m_a \lesssim m_\pi/3$.

Following Ref.~\cite{Bringmann:2006mu} and plugging in the matrix element of Eq.~\eqref{matrix}, the leading order piece of the collision term is then
\begin{align*} 
C^2 &= \frac{\pi \,m_\pi^4}{8 g_\pi (2 \pi)^3 f_{a\pi}^4} \int \frac{\dbar^3 p_{a_1}}{2 E_{a_1}} \int d {\Omega}_2\, d E_{a_2}\, p_{a_2} \Bigg[ \left(\frac{q^2}{E_{\pi_1}^2 T_\pi} - \frac{(\vec{q}\cdot \vec{p}_{\pi_1})^2}{E_{\pi_1}^3 T_\pi^2}- \frac{3(\vec{q}\cdot \vec{p}_{\pi_1})^2}{E_{\pi_1}^4 T_\pi} \right) J' \delta(E_{a_1} - E_{a_2} )\\&
+ \left( \frac{q^2}{E_{\pi_1}^2}J - \frac{3(\vec{q}\cdot \vec{p}_{\pi_1})^2}{E_{\pi_1}^4}J + \frac{2(\vec{q}\cdot \vec{p}_{\pi_1})^2}{E_{\pi_1}^3 T_\pi}J' \right)\partial_{E_{a_2}}\delta(E_{a_1} - E_{a_2}) + \frac{2(\vec{q}\cdot \vec{p}_{\pi_1})^2}{E_{\pi_1}^3}J \,\partial_{E_{a_2}}^2\delta(E_{a_1} - E_{a_2}) \Bigg] \numberthis\\
&= \frac{2 m_\pi^4 e^{-E_{\pi_1}/T_\pi} e^{-m_a/T_a}(T_a -T_\pi)T_a^3(m_a^2 + 3 m_a T_a + 3 T_a^2)}{ g_\pi (2 \pi)^3 f_{a \pi}^4E_{\pi_1}^2 T_a T_\pi} \numberthis \end{align*}
where $J'\equiv e^{-E_{\pi_2}/T_\pi} e^{-E_{a_2}/T_a}$.
This feeds into the calculation of the energy transfer rate of Eq.~\eqref{eq:transferrate} under the assumption that the pions are non-relativistic,
\begin{align*} n_\pi n_a \avg{\sigma v \Delta E} 
 & = \frac{6  e^{-m_\pi/T_\pi} e^{-m_a/T_a}(T_\pi -T_a)T_a^3(m_a^2 + 3 m_a T_a + 3 T_a^2) \sqrt{\pi (m_\pi T_\pi)^5}}{ g_\pi \sqrt{2} (2 \pi)^5 f_{a \pi}^4 T_a T_\pi} \numberthis \\
 &=  \frac{3 m_\pi  e^{-m_a/T_a}(T_\pi -T_a)T_a^2 (m_a^2 + 3 m_a T_a + 3 T_a^2) }{g_\pi (2 \pi)^3 f_{a \pi}^4} \,n_\pi  \end{align*}
By analogy to Eq.~\eqref{eq:equalization}, the term in the Boltzmann equations that equalizes temperatures between the two sectors is \beq \frac{3 g_a^2 m_\pi^3 m_a^2}{(2 \pi)^3 f_{a \pi}^4 H|_{T =m_\pi}} \left(\frac{T_a}{m_\pi}\right)^{-1} \left(\frac{T_\pi}{m_\pi}\right)^{2} \left(\frac{T_\pi-T_a}{m_\pi}\right)e^{-m_a/T_a} \eeq
and decoupling happens when this is order unity. The requirement that thermal decoupling happens after pion freeze-out can be recast as a requirement on $f_{a \pi}$ \beq f_{a \pi}\lesssim \left( \frac{3 g_a^2  m_\pi^3m_a^2 }{ (2 \pi)^3 H|_{T = m_\pi}}  \left(\frac{T_F}{m_\pi}\right)^{2} e^{-m_a/T_F} \right)^{1/4}. \label{answer}\eeq
Eqs.~\eqref{answer} and~\eqref{eq:nonrel_answer} do not match exactly due to a difference in numerical prefactors for BE vs. MB statistics. When that relative factor $\pi^4/90$ is taken into account, then the two match exactly.

\section{axion-SM thermalization}
\label{app2}
In order for the pions to maintain thermal equilibrium with the SM, two conditions must be satisfied: the decays and inverse decays of the axions need to thermalize the axions with the SM, and the axions need to lose kinetic energy via decays faster than they gain energy from the pion $3\to2$ heating. We have verified numerically that as long as these conditions are satisfied at the freeze-out temperature of the pion, the relic abundance of DM is unaffected and the pions constitute cold DM. 

\subsection{Axions in thermal equilibrium with photons}
To understand the requirement that axions maintain thermal contact with the SM, we can ignore the pions and consider only the relevant Boltzmann equations for the axions: \begin{align}
&\frac{\partial n_{a}}{\partial t } + 3 H n_{a} = -\Gamma_a m_a \left( \left< E^{-1}_a\right>_{T_a} n_{a} - \left< E^{-1}_a\right>_{T_\text{SM}} n_a^{\rm eq , T_\text{SM}}\right)\quad\quad\quad \\
&\frac{\partial \rho_{a}}{\partial t} + 3 H \left(\rho_{a} + P_{a}  \right) =  -m_a \Gamma_a \left( n_{a} - n_a^{\rm eq, T_\text{SM} }\right)
\end{align}
where the average axion energy is $ \left< E_a \right> = \rho_a/n_a$. To make the notation less cumbersome for the remainder of this Appendix, the label for equilibrium distributions will denote chemical equilibrium \emph{and} kinetic equilibrium between the axions and SM, \emph{i.e.} $T_a = T_\text{SM} \equiv T$. With this notation, these equations can be re-expressed as
\onecolumngrid
\beq
-T\frac{\partial n_{a}}{\partial T} + 3  n_{a} = -\frac{m_a \Gamma_a}{H} n_{a}  \left( \left< E_a^{-1} \right> - \left< E_a^{-1} \right>^{\rm eq} \frac{n_a^{\rm eq } }{n_{a} }\right) \equiv-\frac{m_a \Gamma_a}{H} n_{a}  c_n \label{eq:firstone} \\
-T \frac{\partial \left< E_a \right> n_{a}}{\partial T} + 3 \left< E_a \right> n_{a} \left(1+w_a  \right) =  -\frac{m_a \Gamma_a}{H} n_{a}   \left( 1 -  \frac{n_a^{\rm eq }}{n_{a} }\right) \equiv  -\frac{m_a \Gamma_a}{H} n_{a} c_\rho, \label{eq:secondone}
\eeq
where $w_a$ is the equation of state of the axion, which is a function of time and axion temperature $w_a=w_a(T_a)$. 
In order to eliminate $\partial n_{a}/\partial T$, Eqs.~\eqref{eq:firstone} and~\eqref{eq:secondone} can be combined to give a differential equation for $ \left< E_a \right>$:
\beq
\frac{\partial \left< E_a \right>}{\partial T} = 3 w_a \frac{ \left< E_a\right>}{ T} - \frac{m_a \Gamma_a}{T H} \left({\left< E_a \right>} c_n - c_\rho \right).
\eeq
In the first term, the expansion is driving the change in the average energy, while in the second term, the decay is driving the average energy. One can check that the first term matches the expectations for a decoupled particle. 
Meanwhile, for further examination of the second term, we define the variable $\alpha=\alpha(T_a)$ such that \beq
\left< E^{-1}_a \right> \equiv \alpha / \left< E_a \right>.
\eeq
The value of $\alpha$ changes monotonically $\alpha \in [1, \pi^6/(360\, \zeta(3)^2) ]$ as $T_a$ goes from 0 to $\infty$. Using this definition, we find
\beq
\frac{\partial \left< E_a \right>}{\partial T} = 3 w_a \frac{ \left< E_a \right>}{ T} + \frac{m_a \Gamma_a}{T H} \left[(1-\alpha) - \frac{n_a^{\rm eq }}{n_{a} }\left(1- \alpha \frac{ \left< E_a \right>}{ \left< E_a\right>^{\rm eq}}\right)   \right]. \label{mastereq}
\eeq
The second term vanishes when the particle is in equilibrium, i.e., $n_a = n_a^{\rm eq }$ with $T_a=T$. If the particle is driven out of equilibrium (for instance by the expansion), this term will push it back into equilibrium. In order to overcome the expansion, $\Gamma_a$ needs to be large enough so that the second term is larger than the first.

First we consider the case that $m_a \ll T$ such that $w_a=1/3$ and $\alpha =\pi^6/(360\, \zeta(3)^2)$. Assuming the axion is near equilibrium and expanding Eq.~\eqref{mastereq} around $T_a = T$ to leading order gives
\beq
\frac{\partial \left< E_a \right>}{\partial T} = \frac{ \left< E_a \right>}{ T} - \frac{m_a \Gamma_a}{T H} (2\alpha-3) \frac{(T_a - T)}{T} .
\eeq
If the two temperatures differ, then the second term will drive the system back into equilibrium if it is comparable to the first,
\beq
\label{eq:lowma_therm}
\frac{m_a \Gamma_a}{T H} \gtrsim \frac{\left<E_a\right>}{(2 \alpha - 3) T} \simeq 4  ~~~~~~~~~~ m_a \ll T.
\eeq

Below, we find the strongest requirement on kinetic equilibrium when $m_a \gtrsim m_\pi$. For intermediate masses $T\lesssim m_a \lesssim m_\pi$, we analytically continue the condition in Eq.~\eqref{eq:lowma_therm} and require
\beq
 \frac{m_a \Gamma_a}{ T H} e^{-m_a/2T} \gtrsim 4 ~~~~~~~~~~ m_a \lesssim m_\pi.
\eeq
We have checked numerically that this requirement ensures thermal equilibrium between the axions and SM for the entire region $m_a \lesssim m_\pi$. This is the condition listed in Eq.~\eqref{eq:aSMkin}.

\subsection{Energy transfer through decays faster than from cannibalization}

The second condition requires that the kinetic energy transferred to the axions from pion $3\to2$ can be compensated by axion decays. This condition will only be important when the axion is heavier than the pion and has a smaller number density. The rate of kinetic energy density loss through decays for non-relativistic axions is $ \Gamma_a n_a T^2/m_a$. Meanwhile, for the axions to sufficiently cool the pions, we require that $\left<\sigma v \Delta E\right> n_a n_\pi \gtrsim m_\pi \dot{n}_\pi \sim H m_\pi^2  n_\pi /T$ at freeze-out when the pions are still barely in chemical equilibrium. Therefore, the requirement is
\beq
 \frac{\Gamma_a T^2 n_a }{m_a} \gtrsim  \frac{H m_\pi^2 n_\pi}{T} 
\eeq 
at freeze-out. We have numerically checked that this condition keeps the pion, axion, and SM at the same temperature in the regime $m_a \gtrsim m_\pi$.

\twocolumngrid

%

\end{document}